\documentclass[prd,a4paper,superscriptaddress,nofootinbib,10pt]{revtex4} 
\usepackage{graphicx}
\usepackage{amssymb}
\usepackage{amsmath}
\usepackage{lscape}
\usepackage{mathrsfs}
\usepackage{textcomp}
\usepackage{epsfig}
\usepackage{slashed}
\usepackage{comment}
\usepackage{xcolor}
\usepackage{enumerate}
\renewcommand{\d}{\mathrm{d}}
\begin{document}
\begin{flushleft}
\texttt{RBI-ThPhys-2022-33}

\texttt{ZTF-EP-22-04}
\end{flushleft}

\title{Noncommutativity and logarithmic correction to the black hole entropy}

\author{Kumar S. Gupta}
\email{kumars.gupta@saha.ac.in}
\affiliation{Theory Division, Saha Institute of Nuclear Physics, 1/AF Bidhannagar, Kolkata 700064, India}

\author{Tajron Juri\'c}
\email{tjuric@irb.hr}

\author{Andjelo Samsarov}
\email{asamsarov@irb.hr}
\affiliation{Rudjer Bo\v{s}kovi\'c Institute, Bijeni\v cka  c.54, HR-10002 Zagreb, Croatia}

\author{Ivica Smoli\'c}
\email{ismolic@phy.hr}
\affiliation{Department of Physics, Faculty of Science, University of Zagreb, 10000 Zagreb, Croatia}
\date{\today}

\begin{abstract}

We study the noncommutative corrections to the entropy of the Reissner-Nordstr\"{o}m black hole using a $\kappa$-deformed scalar probe within the brick-wall framework. The noncommutativity is encoded in an Abelian Drinfeld twist constructed from the Killing vector fields of the  Reissner-Nordstr\"{o}m black hole. We show that the noncommutative effects naturally lead to a logarithmic correction to the Bekenstein-Hawking entropy even at the lowest order of the WKB approximation. In contrast, such logarithmic corrections in the commutative setup appear only after the quantum effects are included through higher order WKB corrections or through higher loop effects. Our analysis thus provides further evidence towards the hypothesis that the noncommutative framework is capable of encoding quantum effects in curved spacetime.
\end{abstract}

\maketitle

\section{Introduction}
Interpretation of the horizon area as the black hole entropy was proposed by Bekenstein \cite{Bekenstein} and Hawking \cite{Hawking} within the framework of semi-classical gravity. Initial attempts to understand the origin of the black hole entropy involved statistical mechanics of the in-falling particles \cite{thorne-zurek} or that of scalar fields, using a brick-wall cutoff to regulate the ultraviolet divergence \cite{thooft,thooft1}. 
 Subsequently, there have been  attempts to obtain the Bekenstein-Hawking area law from various microscopic descriptions of gravity, including string theory \cite{vafa}, quantum geometry \cite{ashtekar} and conformal field theory \cite{strominger, carlip}. This process went beyond the validation of the Bekenstein-Hawking formula and a nonperturbative quantum geometry approach predicted a logarithmic correction to the black hole entropy \cite{kaul}. Such logarithmic corrections have since been found in various quantum descriptions of black holes including conformal field theory \cite{Fursaev:1994ea,carlip1,ksgsen}, string theory \cite{asen1,asen2} and within the context of  AdS/CFT duality \cite{sud}. The higher loop corrections \cite{solo,solo1} as well as nonlocal effective field theories of gravity \cite{xiao, Calmet:2021lny} also lead to such logarithmic corrections. The area law of the black hole entropy can be also related to the entanglement between degrees of freedom inside and outside the horizon \cite{bombelli-sorkin, srednicki} and various corrections to the area law in that context have also been found  in \cite{das}.\\

There is another approach to microscopic description of the spacetime which offers an opportunity to derive the Bekenstein-Hawking area law.
The theory of general relativity together with the quantum uncertainty principle suggests that the spacetime coordinates obey the noncommutative (NC) algebra \cite{dfr1,dfr2}. The NC algebra  comes with a fundamental length parameter, whose effect is expected to be manifest at the quantum gravity scale, or equivalently the Planck scale. Another feature of the NC algebra is that it is inherently non-local. This is made explicit by writing the NC algebra in terms of the star product \cite{szabo}, which often can be defined using the Drinfeld twist operator \cite{Chaichian:2004za} (see also \cite{Aschieri1, Aschieri2}). Any physical theory described within the NC framework is expected to capture certain quantum and non-local effects. In this paper we propose to investigate to what extent such an expectation is actually realized in the context of black hole entropy.\\

One way to analyze how the NC framework captures the quantum effects is to explore the NC corrections to the black hole entropy. To that end, we investigate the entropy of the NC  Reissner-Nordstr\"{o}m (RN) black hole in the brick-wall approach using the WKB technique \cite{thooft,thooft1}. In a commutative theory, the WKB method in the leading order does not generate any logarithmic correction to the black hole entropy. Only after including the higher order WKB corrections the logarithmic terms appear \cite{sub}. This happens as WKB is a semi-classical method and the higher order terms are necessary in the commutative framework to capture the quantum effects. We will show here that in the NC setup, even at the lowest order in WKB, the logarithmic terms appear naturally in the expression of the black hole entropy. This is consistent with and provides further evidence towards the idea that the quantum and nonlocal effects are already built into the NC spacetime algebra \cite{dfr1,dfr2}. \\


In our analysis, we shall use a NC scalar field in a classical RN background, coupled to a  NC $U(1)$ gauge field \cite{andelo, DimitrijevicCiric:2019hqq, Ciric:2019urb, Konjik:2020fle}. The NC algebra used in our model is the $\kappa$-deformed spacetime\footnote{The $\kappa$-deformation in question is defined by the Lie algebra relations
given in \eqref{commDecartes}. This Lie algebra is the Lie algebra of symmetries of a $2D$ plane
$(x,y) =$ semidirect product of $2D$ translations and rotations generated by a $3rd$ axis, $t$. The NC algebra used in \cite{C1,C2} are the simple
Lie algebras $so(3)$ or $so(2,1)$, which by a proper rescaling of generators can
be reduced to the  Lie algebra considered in the present paper.} \cite{C1, C2, L1, L2, L3, L4}, which is well known to be relevant for black hole physics \cite{brian1, brian2, schupp, ksgbh} and cosmology \cite{ohl}. This  NC algebra associated with a particular type of  Drinfeld twist operator, known as the angular twist  \cite{Lizzi, andelo, DimitrijevicCiric:2019hqq}, which respects the symmetries of the RN black hole and can be expressed in terms of 
Poincar\'{e} generators, in contrast to certain previous attempts \cite{Govindarajan:2009wt, Dimitrijevic:2011jg, dim2}. The model used in our analysis has been  derived in \cite{andelo} by first developing a  NC differential calculus
along the lines of \cite{22, 23} and then by using the Seiberg-Witten map \cite{seiberg-witten, seiberg-witten2} to derive the NC gauge theory (see Appendix B for details.) To our knowledge, the analysis presented in this paper provides the first derivation of the logarithmic correction to the entropy of a 3+1 dimensional black hole arising from a Drinfeld twist, which is completely different from similar attempts in the context of dilatonic black holes in 1+1 dimensions \cite{nicolini}, the GUP framework \cite{gup} or noncommutative Schwarzschild black hole \cite{Banerjee:2008gc}. \\

This paper is organized as follows. In section II we first briefly review the main steps of the brick-wall method \cite{thooft,thooft1}  and in IIA we highlight the the role of the near-horizon limit for obtaining the entropy of the the Schwarzschlid black hole within the same model. In section IIB we present the calculation for the entropy of the RN black hole with the charged probe. Section III deals with the NC corrections to the entropy of the RN black hole. Here we introduce the $\kappa$-deformed spacetime noncommutativity and the angular Drinfeld twist to obtain the NC corrections to the RN black hole entropy. Our analysis in this Section demonstrates that for the NC RN case, a logarithmic correction to the black hole entropy arises in the lowest order  in WKB, which is one of the main results of this paper. In Section IV we discuss how to get obtain the entropy in the near Schwarzschild limit as the charge of the RN black hole tends to zero. Section V concludes the paper with some final remarks and an outlook. The paper also contain two appendices to augment the main text.\\

Throughout the paper we are using the natural units where $k_B=\hbar=c=1$.\\

\section{Black hole entropy via the brick-wall model}
It is known that any type of field theoretical calculation of the entropy of a black hole leads to a UV divergence, which requires regularization to extract the relevant physical part \cite{solo1}. The brick-wall method, developed by 't Hooft \cite{thooft,thooft1}  provides an appropriate regularization using the WKB approximation and is compatible with the one-loop calculations \cite{Demers:1995dq}. In this method the analysis starts by examining the Klein-Gordon equation in a fixed black hole background described by the metric $g_{\mu\nu}$
\begin{equation}\label{kg}
\Box_{g}\Phi=0.
\end{equation}
 We separate the Klein-Gordon equation in the time, radial and angular parts using the ansatz $\Phi(x)=e^{-iEt}R_{lm}(r)Y_{lm}(\theta, \phi)$, where $l$ and $m$ denote the angular and azimuthal quantum numbers. Next we use the brick-wall boundary condition given by 
\begin{equation}
R_{lm}(r)=0 \ \text{for} \ r=r_{+}+h \ \text{and}\ \ r=L
\end{equation}
is imposed, where $r_+$ is the radius of the outer horizon, $h$ is the brick-wall cutoff and $L$ is the infrared cutoff. Using the WKB approximation, the radial part of the scalar field can be written as 
\begin{equation}
R_{lm}(r)=e^{i\int{k(r)}dr},
\end{equation}
which is plugged into \eqref{kg} to obtain an equation for the radial wave number
\begin{equation}
k^2=H(r,E,l,m),
\end{equation}
where $H$ is a function of $r$ and other parameters of the theory. As long as $k$ is non-negative the number of radial modes $n$ can be obtained by the Bohr-Sommerfeld rule
\begin{equation}\label{BSrule}
\pi n=\int^{L}_{r_+ +h}k(r,E,l,m)dr,
\end{equation}
which gives the quantization of the energy. The total number of wave solutions with the energy less or equal to $E$ is given by
\begin{equation}\label{number}
N(E)=\sum^{l_{max}}_{l=0}\sum^{l}_{m=-l}n \ \longrightarrow\ \frac{1}{\pi}\int^{L}_{r_+ +h}dr\int^{l_{max}}_{0}dl\int^{l}_{-l}dm\ \ k(r,E,l,m),
\end{equation}
where the summation has been replaced with the integral.
 Notice that $N(E)$ counts the number of classical eigenmodes of a scalar field in the vicinity of a black hole. We are interested in the thermodynamic properties of such a system and assume that each eigenmode may be occupied by any integer number of quanta. The free energy $F$ at some inverse temperature $\beta$ of such a system is given by
\begin{equation}\label{F}
F=-\int^{\infty}_{0}\frac{N(E)}{e^{\beta E}-1}dE=-\frac{1}{\pi}\int^{\infty}_{0}dE\int^{L}_{r_+ +h}dr\int^{l_{max}}_{0}dl\int^{l}_{-l}dm\ \ \frac{k(r,E,l,m)}{e^{\beta E}-1}.
\end{equation}
When calculating \eqref{F} one has to be careful and integrate only over the values for which the radial wave number $k$ is non-negative. We will see later that this means that $l_{max}$ depends on $r$. Also, we are interested in the main contributions to the free energy $F$ coming from the horizon. Therefore the integral over $r$ can be split into two parts
\begin{equation}\label{spliting}
\int^{L}_{r_+ +h}dr(...)= \int^{R}_{r_+ +h}dr(...) +\int^{L}_{R}dr(...),
\end{equation}
where in the first term one keeps the most divergent part as $h\rightarrow 0$, while in the second term $L\gg r_+$. The second term constitutes the usual contribution from the vacuum surrounding the system at large distances and can be omitted \cite{thooft}. Once the free energy $F$ is calculated and the most divergent part in $h$ identified, one can find the entropy of the black hole by simply using
\begin{equation}\label{S0}
S=\beta^2\frac{d F}{d \beta}.
\end{equation}
  Finally one gets rid of the brick-wall cutoff $h$ by equating \eqref{S0} to the Bekenstein-Hawking entropy \cite{Bekenstein, Hawking} and demanding that $\beta$ is  the inverse Hawking temperature.\\
	
	In the following subsection we will outline the brick-wall method for the case of Schwarzschlid and RN black holes in order to fix the notation before introducing the NC effects.

\subsection{Entropy of the Schwarzschlid black hole}

The entropy of the Schwarzschlid black hole in the brick-wall method is well known \cite{thooft}, but here we will repeat the crucial steps in the calculation to emphasize the role of the near horizon region. We start with  the  metric for Schwarzschild black hole
\begin{equation}
ds^2=\left(1-\frac{2GM}{r}\right)dt^2-\left(1-\frac{2GM}{r}\right)^{-1}dr^2-r^2d\Omega^2
\end{equation}
and derive the radial part of the Klein-Gordon equation for a massless scalar as
\begin{equation}
\left(1-\frac{2GM}{r}\right)R^{\prime\prime}_{lm}+\frac{2}{r}\left(1-\frac{GM}{r}\right)R^{\prime}_{lm}-\left[\frac{l(l+1)}{r^2}-\frac{E^2}{1-\frac{2GM}{r}}\right]R_{lm}=0.
\end{equation}
Next we use the WKB approximation and get the radial wave number $k_S (r,l,E)$
\begin{equation}
k_{S}^{2}=\frac{1}{1-\frac{2GM}{r}}\left(\frac{E^2}{1-\frac{2GM}{r}}-\frac{l(l+1)}{r^2}\right)
\end{equation}
and the number of radial modes $n$ is given by the Bohr-Sommerfeld rule \eqref{BSrule}.
We keep $k_S$ non-negative which fixes the maximal value of the orbital quantum number as
\begin{equation}
l_{max}(l_{max}+1)=\frac{E^2r^2}{1-\frac{2GM}{r}}.
\end{equation} 
We define a new coordinate as $x=r-2GM$.
The total number $N$ of wave solutions \eqref{number}  is given by
\begin{equation}
N_S(E)=\frac{1}{\pi}\int^{L-2GM}_{h}dx\frac{x+2GM}{x}\int^{l_{max}}_{0}dl\ \ (2l+1)\sqrt{E^2-\frac{x}{x+2GM}\frac{l(l+1)}{(x+2GM)^2}}.
\end{equation}
We now look for the main contributions and split the integral as in \eqref{spliting} to obtain
\begin{equation}\label{Nsch}
N_S(E)=\frac{32G^4M^4E^3}{3\pi h}+\frac{E^3L^3}{\pi}.
\end{equation}
Notice that when evaluating the first term in \eqref{spliting}  the maximal value of $l$ is calculated in the near horizon limit with $x\approx h$ and gives $l_{max}(l_{max}+1)=\frac{(2GM)^3E^2}{x}$. While evaluating the second term we are in the $x\gg 2GM$ limit which gives $l_{max}(l_{max}+1)=E^2L^2$. The free energy is given by
\begin{equation}\label{Fsch}
F_S=-\frac{2\pi^3}{45h}\left(\frac{2GM}{\beta}\right)^4 - \frac{2L^3\pi^3}{30\beta^4}.
\end{equation}
The second term in \eqref{Fsch} is the contribution from the vacuum surrounding the system at large distances and can be omitted, while the first term is the intrinsic contribution from the horizon that diverges as $h\rightarrow 0$. The contribution of the horizon to the entropy  is
\begin{equation}
S_{S}=\frac{8\pi^3}{45h}2GM\left(\frac{2GM}{\beta}\right)^3
\end{equation}
which is in complete agreement with \cite{thooft}.
Now one can obtain the value for the cut-off $h$ by imposing that the temperature is the Hawking temperature $T_H=\frac{1}{\beta}=\frac{1}{8\pi MG}$ and entropy is the Bekenstein-Hawking entropy $S=\frac{A}{4G}=4\pi GM^2$
\begin{equation}\label{hsch}
h=\frac{1}{720\pi M}
\end{equation}
which is in agreement with \cite{thooft}. Notice that even though eq. \eqref{hsch} might suggest that the brick-wall cutoff depends on $M$, it is actually a coordinate artifact since calculating the invariant distance shows that the brick-wall can be seen as a property of the horizon that is independent of the particular size of the black hole in question.

\subsection{Entropy of the Reissner-Nordstr\"om black hole}
The RN black hole is a  solution of the coupled Einstein-Maxwell system of equations for both the metric $g_{\mu\nu}$ and vector potential $A_{\mu}$. The RN metric is given by
\begin{equation}
ds^2=fdt^2-f^{-1}dr^2-r^2d\Omega^2, \quad f=1-\frac{2GM}{r}+\frac{Q^2 G}{r^2}.
\end{equation}
The Klein-Gordon operator in the RN background is obtained by the minimal substitution $$\Box_g=g^{\mu\nu}\nabla_{\mu}\nabla_{\nu} \longrightarrow g^{\mu\nu}D_{\mu} D_{\nu},$$ where $D_{\mu}=\nabla_\mu-iqA_{\mu}$ and $A_{\mu}=-\delta^{t}_{\mu}\frac{Q}{r}$ is the fixed background electromagnetic potential of the  RN black hole. The radial part of the  Klein-Gordon equation for a massless minimally coupled scalar field of charge $q$ in the RN background can be written as
\begin{equation}
fR^{\prime\prime}_{lm}+\frac{2}{r}\left(1-\frac{GM}{r}\right)R^{\prime}_{lm}-\left[\frac{l(l+1)}{r^2}-\frac{1}{f}\left(E-\frac{qQ}{r}\right)^2\right]R_{lm}=0.
\end{equation}
We use the WKB approximation and get the radial wave number $k_{RN}(r,l,E)$
\begin{equation}
\label{krn}
k_{RN}^{2}=\frac{1}{f}\left[\frac{1}{f}\left(E-\frac{qQ}{r}\right)^2-\frac{l(l+1)}{r^2}\right].
\end{equation}
The number of radial modes $n$ is given by the Bohr-Sommerfeld rule \eqref{BSrule}
where we have to keep $k_{RN}$ non-negative, so the maximal value of the orbital quantum number is 
\begin{equation}
l_{max}(l_{max}+1)=\frac{r^2}{f}\left(E-\frac{qQ}{r}\right)^2.
\end{equation}
 The total number $N_{RN}$ of wave solutions \eqref{number}
after switching to near horizon coordinates $x=r-r_+$ is given by
\begin{equation}\label{Nrn}
N_{RN}(E)=\frac{1}{\pi}\int^{L-r_+}_{h}dx\frac{x+r_+}{\sqrt{x(x+r_+ -r_{-})}}\int^{l_{max}}_{0}(2l+1)dl \left[\frac{\left(E-\frac{qQ}{x+r_+}\right)^2(x+r_+)^2}{x(x+r_+ -r_{-})}-\frac{l(l+1)}{(x+r_+)^2}\right]^{\frac{1}{2}}
\end{equation}
where we used
\begin{equation}
r_{\pm}=GM \pm \sqrt{G^2M^2-GQ^2}.
\end{equation}
We can split the integral as in (\ref{spliting}) and look for the intrinsic contribution from the near horizon region  when $h\rightarrow 0$ and omit the contribution for the surrounding vacuum. Since the integrals over $l$ and $x$ are of the same type as in the Schwarzschild case it is easy to see that the near horizon contribution is given by
\begin{equation}\label{Nrnn}
N_{RN}(E)=\frac{2}{3\pi}\frac{r^{6}_{+}\left(E-\frac{qQ}{r_+}\right)^3}{(r_+ - r_-)^2}\frac{1}{h}
\end{equation}
where we also used
\begin{equation}
 l_{max}(l_{max}+1)=\frac{r^{4}_+}{x(r_+-r_-)}\left(E-\frac{qQ}{r_+}\right)^2.
\end{equation}
The free energy is given by
\begin{equation}\label{Frn}
F_{RN}=-\frac{2}{3\pi}\frac{r^{6}_{+}}{(r_+ - r_-)^2}\frac{1}{h}K(\beta), \quad K(\beta)=\int^{\infty}_{0}dE\frac{\left(E-\frac{qQ}{r_+}\right)^3}{e^{\beta E}-1}.
\end{equation}
Notice that  $K(\beta)$ has an infinite contribution coming from the electrostatic self-energy of charge $q$ of the scalar particle. This contribution vanishes when $q\rightarrow 0$.  Using the $\zeta$-function regularization $K(\beta)$ can be written as
\begin{equation}
K(\beta)=\frac{\Gamma(4)\zeta(4)}{\beta^4}-\frac{3qQ\Gamma(3)\zeta(3)}{r_+ \beta^3}+\frac{3q^2Q^2\Gamma(2)\zeta(2)}{r^{2}_{+}\beta^2}-\frac{q^3Q^3\Gamma(1)\zeta(1)}{r^{3}_{+}\beta},
\end{equation}
where the infinite contribution appears in $\zeta(1)$. In subsequent analysis we shall ignore this infinite contribution from the electrostatic self-energy, whereby the entropy is given by
\begin{equation}\label{Srn}
S_{RN}=-\frac{2}{3\pi}\frac{r^{6}_{+}}{(r_+ - r_-)^2}\frac{1}{h}\left(-\frac{4\Gamma(4)\zeta(4)}{\beta^3}+\frac{9qQ\Gamma(3)\zeta(3)}{r_+ \beta^2}-\frac{6q^2Q^2\Gamma(2)\zeta(2)}{r^{2}_{+}\beta}\right),
\end{equation}
The value for the cutoff $h$ is obtained by imposing that the temperature is the Hawking temperature $T_H=\frac{1}{\beta}=\frac{r_+-r_-}{4\pi r^{2}_{+}}$ and entropy is the Bekenstein-Hawking entropy $S=\frac{A_{RN}}{4G}=\frac{\pi r^{2}_{+}}{G}$, which gives
\begin{equation}\label{hrn}
h=\frac{G}{360\pi }\frac{r_+-r_-}{ r^{2}_{+}}\left(1-\frac{270qQ\zeta(3)}{\pi^3}\frac{r_+}{r_{+}-r_-}+60q^2Q^2\frac{r^{2}_{+}}{(r_{+}-r_-)^2}\right).
\end{equation}
For $Q\rightarrow 0$ this is in agreement with (\ref{hsch}) and this recovers the entropy for the Schwarzschild black hole. We can check that
\begin{equation}
 \lim_{q\longrightarrow 0}S_{RN}=\frac{8\pi^3}{45}\frac{r^{6}_{+}}{(r_+ -r_-)^2}\frac{1}{h\beta^3}
\end{equation}
which is in agreement with \cite{Demers:1995dq, inside}.

\section{Noncommutative correction to the entropy of the Reissner-Nordstr\"om black hole}\label{sec3}

In this paper we shall consider a $\kappa$-deformed  spacetime noncommutativity, whose commutation relations in the Cartesian coordinates are given by
\begin{equation}\label{commDecartes}
[t, x]_{\star}  =  - ia y , \quad [t, y]_{\star}  =  ia x, \quad [x, y]_{\star}  =  0,
\end{equation}
where $a$ is the NC deformation parameter. The corresponding relations in the spherical coordinates can be written as
\begin{equation}\label{commrelvarious}
 [r,  t]_{\star} =  0, \quad  [r,  e^{i\phi}]_{\star} =  0, \quad [ t, e^{i\phi} ]_{\star}  =  -a e^{i\phi}.
\end{equation}
The star product of two functions $f$ and $g$ appearing in (\ref{commDecartes}) and (\ref{commrelvarious}) is defined as 
\begin{eqnarray}
f\star g =  \mu \{ \mathcal{F}^{-1} f\otimes g \} \label{fStarg0Phi},
\end{eqnarray}
where $\mu$ is the usual point-wise multiplication of functions, $\mu (f \otimes g) = f \cdot g$ and $\mathcal{F}$ is the twist operator given by  \cite{Lizzi, andelo, DimitrijevicCiric:2019hqq}  
\begin{equation}
\mathcal{F}   \equiv e^{-\frac{i}{2}\theta ^{\alpha\beta}\partial_\alpha\otimes \partial_\beta} = e^{-\frac{ia}{2} (\partial_t\otimes\partial_\phi - \partial_\phi\otimes\partial_t)},
\label{AngTwist0Phi}
\end{equation}
where $\alpha,\beta = t,r, \theta, \phi$ and $\theta^{t\phi}= -\theta^{\phi t}=a$  are the only non-zero components of the deformation tensor $\theta^{\alpha \beta}$.  This twist operator is formed from the Killing vector fields of the RN metric. In addition,
$X_{1}=\partial _t$,
$X_{2}= \partial_\phi$ are commuting vector fields,
$[X_{1},X_{2}]=0$, rendering  (\ref{AngTwist0Phi}) to be an Abelian twist. We call 
(\ref{AngTwist0Phi}) an angular twist because the vector field $X_2 =\partial_\varphi = x\partial_y - y\partial_x$ is 
a generator of rotations around the $z$-axis. Since the twist (\ref{AngTwist0Phi}) is defined in terms of the  Killing vector fields of the RN metric, it
 leaves the RN metric unchanged.
An important aspect of the twist  (\ref{AngTwist0Phi}) is that it is expressed entirely in terms of Poincar\'e generators, unlike some previous attempts to construct NC gauge theory \cite{Govindarajan:2009wt, Dimitrijevic:2011jg, dim2}, where the twist involves the generators of ${\it{igl}} (1,3),$ 
The twisted deformation considered here is a special case of deformations introduced in \cite{luk1,luk2}. Similar type of twist operators that lead to a Lie algebra-type of  deformation of Minkowski space-time were also considered in \cite{Meljanac:2017oek}.

Following \cite{andelo}, we consider a NC charged scalar field in the background of a RN black hole, where the noncommutativity is defined by star product (\ref{fStarg0Phi}) and the angular twist (\ref{AngTwist0Phi}). We refer to this system as NCRN. Up to the first order in the NC deformation parameter $a$, the radial part of the corresponding Klein-Gordon equation is given by (See Appendix B for details)
\begin{equation}    \label{NCRNequation}
fR^{\prime\prime}_{lm}+\frac{2}{r}\left(1-\frac{GM}{r}\right)R^{\prime}_{lm}-\left[\frac{l(l+1)}{r^2}-\frac{1}{f}\left(E-\frac{qQ}{r}\right)^2\right]R_{lm}-ima\frac{qQ}{r^3}\left[\left(\frac{GM}{r}-\frac{GQ^2}{r^2}\right)R_{lm}+rfR^{\prime}_{lm}\right]=0.
\end{equation}
Next we use the WKB approximation and get the radial wave number $k_{NCRN}(r,l,m,E)$ as
\begin{equation}\label{knc}
k_{NCRN}^{2}=\frac{1}{f}\left[\frac{1}{f}\left(E-\frac{qQ}{r}\right)^2-\frac{l(l+1)}{r^2}-ima\frac{qQ}{r^3}\left(\frac{GM}{r}-\frac{GQ^2}{r^2}\right)\right].
\end{equation}
Note that unlike the commutative case,  there is an explicit dependence of $k_{NCRN}$ on the magnetic quantum number $m$ and appearance of the imaginary factor in it. At a  perturbative level,  expanding \eqref{knc} in the NC deformation parameter $a$ we have
\begin{equation}
k_{NCRN}=\frac{1}{\sqrt{f}}(A+aB)^{\frac{1}{2}}=\sqrt{\frac{A}{f}}+\frac{a}{2}\frac{1}{\sqrt{fA}}B-\frac{a^2}{8}\frac{1}{\sqrt{fA^3}}B^2+\mathcal{O}(a^3).
\end{equation}
The first term is exactly the radial wave number in the commutative case  (\ref{krn}), 
while the second and third are NC corrections. Notice that $B$ is linear in the magnetic quantum number $m$. Hence there is no contribution to $N(E)$ from the term linear in $a$  since
\begin{equation}
\sum^{l}_{m=-l}m=0=\int^{l}_{-l}mdm.
\end{equation}
Thus the leading NC correction is quadratic in $a$. Furthermore, all correction that are odd powers in $a$ vanish which means that imaginary part of $k$ does not contribute to the number of solutions $N$. This result can be proven nonperturbatively to all orders in the deformation parameter $a$. In the Appendix A we show that the imaginary part of the wave number $k$ is zero after we integrate over $m$ so it does not contribute to the free energy and entropy in the end. Therefore we can immediately go to the near horizon limit and omit the surrounding vacuum contributions and we get
\begin{equation}
k_{NCRN}=k_{RN}+\tilde{k}, 
\end{equation}
where
 \begin{equation}
k_{RN}=\frac{r_+}{\sqrt{x(r_+ -r_{-})}}\left[\frac{\left(E-\frac{qQ}{r_+}\right)^2 r^{2}_{+}}{x(r_+ -r_{-})}-\frac{l(l+1)}{r^{2}_{+}}\right]^{\frac{1}{2}}, \quad \tilde{k}=\frac{a^2}{8}\frac{x(r_+-r_-)}{r^{2}_{+}\left(E-\frac{qQ}{r_+}\right)^3}\frac{q^2Q^2}{r^{6}_{+}}\left(\frac{GM}{r_+}-\frac{GQ^2}{r^{2}_+}\right)^2 m^2.
\end{equation}
This gives 
\begin{equation}
N_{NCRN}(E)=N_{RN}(E)+\tilde{N}(E),  \quad \tilde{N}(E)=\frac{a^2q^2Q^2}{48\pi}\left(\frac{GM}{r_+}-\frac{GQ^2}{r^{2}_+}\right)^2\frac{E-\frac{qQ}{r_+}}{r_+-r_-}\text{ln}\left(\frac{l_p}{h}\right),
\end{equation}
where $N_{RN}$ is given in \eqref{Nrnn}, $l_p=\sqrt{G}$ is the Planck length and $$l_{max}(l_{max}+1)=\frac{r^{4}_{+}}{r_+-r_-}\frac{\left(E-\frac{qQ}{r_+}\right)^2}{x},$$ which ensures that $k_{RN}$ is nonnegative.
The free energy is given by
\begin{equation}     \label{freeen}
F_{NCRN}=F_{RN}+\tilde{F},  \quad \tilde{F}=-\frac{a^2q^2Q^2}{48\pi}\left(\frac{GM}{r_+}-\frac{GQ^2}{r^{2}_+}\right)^2\frac{1}{r_+-r_-}\text{ln}\left(\frac{l_p}{h}\right)\left(\frac{\pi^2}{6\beta^2}-\frac{qQ}{r_+ \beta}\zeta(1)\right),
\end{equation}
where $F_{RN}$ is given in \eqref{Frn}.
The entropy can be expressed as 
\begin{equation}\label{to}
S_{NCRN}=S_{RN}+\frac{a^2q^2Q^2}{48\pi}\left(1 - \frac{GM}{r_+}\right)^2\frac{1}{r_+-r_-}\text{ln}\left(\frac{l_p}{h}\right)\frac{\pi^2}{3\beta}
\end{equation}
where $S_{RN}$ is given in \eqref{Srn} and we have removed the infinite contribution arising from the electrostatic self-energy like before. 
The cutoff $h$ is fixed by matching $S_{RN}$ with the Bekenstein-Hawking entropy, which is exactly \eqref{hrn} and using this we obtain the final expression of the entropy as
\begin{equation}\label{Sncrn}
S_{NCRN}=\frac{A_{RN}}{4G}+a^2\mathcal{W}\text{ln}\left( \frac{A_{RN}}{l_p}\right)+a^2\mathcal{V}
\end{equation}
where $A_{RN}=4\pi r^{2}_{+}$ is the area of the RN black hole, $\mathcal{W}$ and $\mathcal{V}$ are functions of $r_{\pm}$ and $q$ and can be determined from \eqref{hrn} and \eqref{to}. It is important to note that the NC corrections to the entropy are  similar to the usual subleading quantum corrections (see \cite{sub}) indicating  the nonlocal and quantum nature of the NC algebra all ready in the lowest order in the WKB.

\section{The almost Schwarzschild limit}

The Abelian Drinfeld twist used in this paper does not deform either the Schwarzschild metric or its coupling to the scalar probe. Hence taking limits of $Q,q\longrightarrow 0$  in (\ref{Sncrn}) simply reproduces the commutative results. The noncommutative corrections only appear if we have a charged black hole (in our case RN) and a charged scalar probe. However we can look at the RN black holes with very small charge $Q$ in order to compare the NC corrections with respect to the (almost) Schwarzschild black hole. Therefore, we will look at \eqref{Sncrn} in the limit $Q\longrightarrow 0$, that is we expand everything to the lowest order in the black hole charge $Q$. Since the noncommutative correction is propositional to $Q^2$ it is enough to expand the $S_{RN}$ up to quadratic terms in $Q$ and all the rest keep in the zeroth order. In doing this we used
\begin{equation}
S_{RN}=S_{S}-2\pi\frac{Q^2}{M}+\mathcal{O}(Q^4), \quad h=\frac{1}{720\pi M}+\mathcal{O}(Q^2), \quad \frac{1}{\beta}=\frac{r_+ - r_-}{4\pi r^{2}_{+}}=\frac{1}{8\pi GM}+\mathcal{O}(Q^2)
\end{equation}
and therefore the entropy in the almost Schwarzschild limit has the following form
\begin{equation}
S_{NCRN}=S_{S}-\frac{2\pi Q^2}{M}+\frac{a^2q^2Q^2}{9216 G^2M^2}\text{ln}\left(720\pi Ml_p\right)
\end{equation}
or if we exploit the formula for $S_{S}$
\begin{equation}
M=\sqrt{\frac{S_{S}}{4\pi G}}
\end{equation}
for the $Q\approx 0$ we can obtain
\begin{equation}
S_{NCRN}=\frac{A}{4G}+\mathcal{G}(A)+\mathcal{H}(A) \text{ln}\left(\frac{A}{l^{2}_p}\right)
\end{equation}
where 
\begin{equation}
\mathcal{G}(A)=-(4\pi)^{3/2}\frac{Q^2 l^{2}_{p}}{\sqrt{A}}, \quad \mathcal{H}(A)=a^2\frac{\pi}{1152}\frac{q^2 Q^2}{A}
\end{equation}
and $A=16\pi G^2M^2$ is the area of the Schwarzschild black hole. Note that this result is in a form compatible with the generic form of WKB expansion of the entropy of black holes \cite{sub}, but the main difference is that in the commutative case the logarithmic corrections appear as subleading corrections, while in the NC framework they come in the lowest WKB order.

\section{Final Remarks}

In this paper we have shown that a NC framework can give rise to a
logarithmic correction to the black hole entropy, which is a purely
quantum effect. Our model consists of a $\kappa$-deformed NC scalar field on a classical
RN black hole background, coupled to a NC $U(1)$ gauge field. The NC
algebra considered here arises from an Abelian Drinfeld twist, known as
the angular twist. It is adapted to the isometries of the background RN geometry,
with the associated stationary Killing vector field $\partial_t$ and the axial Killing vector field $\partial_{\phi}$. The analysis was performed using the brick-wall method in the lowest order of the semiclassical WKB approximation. \\

It is well known that in the commutative setup, the lowest order in WKB
does not lead to logarithmic correction to the black hole entropy. There one has to go beyond the semi-classical limit in order to obtain such logarithmic corrections. 
However, the results obtained here indicate that the NC framework is
capable of revealing the quantum effects in the black hole entropy even at the lowest order in the WKB. This is consistent with and provides further evidence towards the hypothesis that the NC framework is capable of encoding quantum effects of curved spacetime \cite{dfr1, dfr2}.\\

In addition, to the best of our knowledge, this work provides the first derivation of
the logarithmic  correction to the Bekenstein-Hawking using a Drinfeld twist arising from a $\kappa$-deformed Hopf algebra. The choice of the
twist operator is certainly not unique, but the angular twist used here
captures a lot of features present in other Abelian twists. It is important to note that the angular twist \eqref{AngTwist0Phi} used in the paper is made of vector fields $\partial_t$ and $\partial_\phi$. As a result, the deformations of any metric  that has these vectors as Killing vector fields are exactly zero. We have considered RN metric in this paper, which has $\partial_t$ and $\partial_\phi$ as Killing vector fields. Consequently, as has been shown in \cite{andelo, DimitrijevicCiric:2019hqq}, the only NC correction
that shows up is the correction between the coupling between the U(1) potential and the charged scalar, leading to \eqref{NCRNequation}. In order to investigate the possibility of  NC corrections to the metric we either have to consider a more general twist \cite{Aschieri:2009qh, Schenkel:2011biz} (not made entirely of Killing vector fields) or  different metrics (with less symmetries). In such cases it is expected in general that there might be additional corrections to the BH entropy, which is a matter of
further analysis.\\

Black hole entropy can be described in the brick-wall model \cite{thooft, thooft1} as well as using entanglement of the degrees of freedom between the two sides of the horizon \cite{bombelli-sorkin, srednicki}. It is a remarkable fact that both these approaches lead to an almost identical UV divergence in the black hole entropy, and it has been argued that the black hole entropy obtained from these seemingly different approaches are indeed related \cite{solo1,das}. The universal nature of such an UV divergence in the black hole entanglement entropy is related to the Type-III nature of the associated von Neumann algebra of observables appearing in these quantum field theories \cite{Witten:2021unn, witten1, witten-longo}.\\

It is natural to inquire if such an universal divergence in the black hole entropy continues to persist within the NC framework. The analysis here indicates that at least in the perturbative limit of a small NC deformation parameter the UV divergence of the black hole entropy persists within the NC framework. This is further supported by similar results for the case of NC BTZ coming from $\kappa$-Minkowski spacetime algebra and from the evaluation of the renormalized entanglement entropy  \cite{Juric:2016zey} using the heat kernel and effective action method \cite{Susskind:1994sm, Fursaev:1994in}. The universal appearance of the UV divergence in these systems suggests that the typology of the associated von Neumann algebra of observables remains unchanged at least when the NC effects are treated perturbatively. Whether there is any change in the typology of the von Neumann algebras when the NC effects are considered nonperturbatively is a much deeper question, which is beyond the scope of the present analysis.

\bigskip
\bigskip

\noindent{\bf Acknowledgment}\\
The authors would like to thank   M. Dimitrijevi\'{c}-\'{C}iri\'{c} and N.Konjik for various discussions.
 This research has been supported by Croatian Science Foundation project IP-2020-02-9614.

\bigskip
\bigskip

\appendix

\section{Nonperturbative integration over $m$}

Let us examine the wave number \eqref{knc} and rewrite it as (up to a sign)
\begin{equation}
k=\left(\alpha-ia\beta m\right)^{1/2}
\end{equation}
where 
\begin{equation}
\alpha=\frac{1}{\sqrt{f}}\left[\frac{1}{f}\left(E-\frac{qQ}{r}\right)^2-\frac{l(l+1)}{r^2}\right], \quad \beta=\frac{qQ}{\sqrt{f}r^3}\left(\frac{GM}{r}-\frac{GQ^2}{r^2}\right).
\end{equation}
A potential problem with this is that $k$ in principle has  real and imaginary parts. We shall show below that the imaginary part does not cause problems since it does not contribute to the free energy and entropy. In order to see that one needs to calculate the number of wave solutions \eqref{number}. In doing so, one needs to perform the integration over the magnetic quantum number $m$ first. Let us therefore consider the  quantity
\begin{equation}
I=\int^{l}_{-l}k dm=\int^{l}_{-l}dm \left(\alpha-ia\beta m\right)^{1/2}
\end{equation}
where $l\in\mathbb{N}$ is the orbital angular momentum. After performing this integral and using the abbreviation $R=\sqrt{\alpha^2+a^2\beta^2l^2}$ and $\tan\chi=-\frac{a\beta l}{\alpha}$ we get
\begin{equation}
I=-\frac{4R^{3/2}}{3a\beta}\sin{\frac{3\chi}{2}}
\end{equation}
which is manifestly real in all orders in $a$. Furthermore we can expand $I$ in the deformation parameter $a$ 
\begin{equation}
I=\frac{4l\sqrt{\alpha}}{3}-a^2\frac{\beta^2l^3}{\alpha^{3/2}}+\mathcal{O}(a^4)
\end{equation}
showing the absence of the linear term in $a$ which is in agreement with the perturbative analysis in Sec.\ref{sec3}.


\section{Some elements of the twist deformation and derivation of  the equation of motion (\ref{NCRNequation})}

In this paper we work with $\kappa$-deformed spacetime noncommutativity for which the defining relations, the star product and the angular twist were given at the beginning of Section III. The twist (\ref{AngTwist0Phi}) may also be written in a Sweedler notation as
\begin{equation}
\mathcal{F} = f^{\alpha} \otimes f_{\alpha},     \quad   {\mathcal{F}}^{-1} = {\bar{f}}^{\alpha} \otimes  {\bar{f}}_{\alpha},
\end{equation}
where for each value of $\alpha,$   $f^{\alpha}, f_{\alpha}, {\bar{f}}^{\alpha} $ and  $  {\bar{f}}_{\alpha}$ are generally all different elements of the universal enveloping algebra of the symmetry algebra in question (in the current context the  Poincar\'e algebra). In view of this, the star product between two functions may be written as $f \star g = \mu \circ {\mathcal{F}}^{-1} (f \otimes g) = {\bar{f}}^{\alpha} (f) \cdot  {\bar{f}}_{\alpha} (g).$

  The twist (\ref{AngTwist0Phi}) satisfies the cocycle and counital conditions:
  \begin{eqnarray} \label{cocycle}
      ({\mathcal{F}} \otimes \mathbf{1} )\cdot ( \triangle \otimes id){\mathcal{F}} & =&
      (\mathbf{1} \otimes {\mathcal{F}}))\cdot ( id \otimes \triangle ){\mathcal{F}}, \nonumber \\
       \mu \circ (\epsilon \otimes id){\mathcal{F}} &= &\mathbf{1} = \mu \circ (id \otimes \epsilon){\mathcal{F}},
  \end{eqnarray}
together with the requirement ${\mathcal{F}} = \mathbf{1}  \otimes  \mathbf{1} +  {\mathcal{O}}(a),$ which ensures  that in the limit of vanishing  deformation, $a \longrightarrow 0,$ an undeformed symmetry algebra is restored. Therefore, it is a Drinfeld twist, ensuring that a twist deformation of the initial symmetry algebra, which is a Hopf algebra, gives rise to a deformed algebraic structure  which is again a Hopf algebra. The deformation itself is carried out by the following set of similarity transformations:
\begin{equation} \label{twisting}
\Delta^{\mathcal{F}} (X)=\mathcal{F}\Delta (X)\mathcal{F}^{-1},
\end{equation}
\begin{equation}
S^{\mathcal{F}}(X)=\chi S(X) \chi^{-1},   
\quad  \epsilon^{\mathcal{F}}(X)=0,
\end{equation}
applied to the generators of the initial Hopf algebra and its structural maps, the coproduct $\Delta,$ the counit $\epsilon$ and the antipode $S.$
In the above relations $\chi =  f^{\alpha} S( f_{\alpha}), ~$  and    $~   {\chi}^{-1} =  S({\bar{f}}^{\alpha})   {\bar{f}}_{\alpha}$.
Note that the $\star $-product (\ref{fStarg0Phi}) is  noncommutative and in the limit $a\rightarrow 0$ of vanishing deformation it
reduces to the usual point-wise multiplication. However, it is also associative and  this property is guaranteed by the first relation in (\ref{cocycle}).
In this way  the noncommutative algebra
of functions, i.e. the noncommutative spacetime comes to light.


The twisted symmetry of the NC
Minkowski spacetime (\ref{commDecartes}) ensuing from the twist (\ref{AngTwist0Phi}) is the twisted Poincar\'e symmetry.
The latter  is described by  the twisted Poincar\'e Hopf algebra whose algebraic and coalgebraic sector were presented in \cite{andelo}.
The  differential calculus appropriate for the above context has been developed from an ordinary differential calculus through a deformation by means of the angular twist operator.  
The details of this can also be found in \cite{andelo, DimitrijevicCiric:2019hqq, Ciric:2019urb, Konjik:2020fle}.


In order to obtain the equation of motion (\ref{NCRNequation}) that is a central point of our study, we start with
 the (RN) metric representing a charged non-rotating black hole  with  mass $M$  and charge $Q$.  We take this solution to represent our nondynamical gravitational background characteristic of the noncommutative semiclassical hybrid model studied in \cite{andelo}.
Being static and spherically symmetric, the  spacetime of RN black hole
 has four Killing vectors, among which  $\partial_t$ and $\partial_\phi $ are included,  and $t$ and $\phi$ are the time and polar variables  
   of the spherical  coordinate system $x^{\mu} = (t,r,\theta,\phi)$. Note that these are exactly the vector fields utilized to build the Abelian Drinfeld twist operator (\ref{AngTwist0Phi}).

 This setting was further used in  \cite{andelo}  to construct a semiclassical model describing  a   charged NC scalar field $\hat{\Phi}$ and  NC $U(1)$ gauge field  $\hat{A}$, both being in a mutual interaction and in an interaction with a classical gravitational background of RN type. The model was built by using deformation quantization techniques as applied to the Drinfeld twist operator (\ref{AngTwist0Phi}) \cite{andelo, DimitrijevicCiric:2019hqq}.
As already pointed out, semiclassical here means that the gravitational field is undeformed by noncommutativity, and   the only degrees of freedom that are  actually deformed  are    the scalar  and   gauge field that propagate in that classical gravitational background.
In a sense, we are thus dealing with a situation where the scalar and gauge field get quantized, while the gravitational field does not. It is however important to stress that the gauge and scalar field are  not quantized in a sense of quantum field theory quantization.

  The action  is given by
\begin{eqnarray}  
S[\hat{\Phi}, \hat{A}] &=& S_\Phi + S_A, \nonumber \\ 
S_\Phi &=& \int \d ^4x \, \sqrt{-g}\star\Big( g^{\mu\nu}\star D_{\mu}\hat{\Phi}^+ \star
D_{\nu}\hat{\Phi} - \mu^2\hat{\Phi}^+ \star\hat{\Phi}\Big) , \label{SPhi}\\
S_A &=& -\frac{1}{4q^2} \int \d^4 x\,\sqrt{-g}\star g^{\alpha\beta}\star g^{\mu\nu}\star
\hat{F}_{\alpha\mu}\star \hat{F}_{\beta\nu} . \label{SA}
\end{eqnarray}
where
\begin{equation}  \label{FF}
\hat{F}_{\mu\nu} = \partial_\mu \hat{A}_\nu - \partial_\nu \hat{A}_\mu -i  \big( \hat{A}_\mu \star \hat{A}_\nu - \hat{A}_\nu \star \hat{A}_\mu \big).
\end{equation}

The  operator $D_{\mu}$ appearing in  (\ref{SPhi}) is the covariant   derivative  of the scalar field $\hat{\Phi}$ and it  is defined as
\begin{equation}
D_\mu\hat{\Phi} = \partial_\mu\hat{\Phi} - i \hat{A}_\mu\star \hat{\Phi} \label{DPhi} . \nonumber
\end{equation}
A favorable  feature is that  due to the twist operator  (\ref{AngTwist0Phi})  not
acting on the metric tensor $g_{\mu \nu}$,   $\star$-products in $\sqrt{-g}\star
g^{\alpha\beta}\star g^{\mu\nu}$ can all be removed.

As pointed out in \cite{andelo},  the functionals (\ref{SPhi}) and (\ref{SA}) are invariant under the following infinitesimal
$U(1)_\star$
gauge transformations:
\begin{eqnarray}
\delta_\star \hat{\Phi} &=& i\hat{\Lambda} \star \hat{\Phi}, \nonumber  \\
\delta_\star \hat{A}_\mu &=& \partial_\mu\hat{\Lambda} + i  \big( \hat{\Lambda} \star \hat{A}_\mu - \hat{A}_\mu \star \hat{\Lambda}  \big),
\label{NCGaugeTransf}\\
\delta_\star \hat{F}_{\mu\nu} &=& i  \big(  \hat{\Lambda} \star  \hat{F}_{\mu\nu} -  \hat{F}_{\mu\nu} \star   \hat{\Lambda}  \big),  \nonumber \\
\delta_\star g_{\mu\nu} &=& 0.\nonumber
\end{eqnarray}
where $\hat{\Lambda}$ is  the NC gauge parameter .

Indeed, the functionals (\ref{SPhi}) and (\ref{SA}) are also invariant under the
finite NC $U(1)_\star $  transformations defined as:
\begin{eqnarray}
\hat{\Phi}' &=& U_\star\star \hat{\Phi}, \nonumber \\
\hat{A}'_\mu &=& -U_\star \star\partial_\mu U_\star^{-1} + U_\star \star \hat{A}_\mu \star U_\star^{-1} ,\nonumber \\
\end{eqnarray}
with 
$ U_\star = e^{i\hat{\Lambda}}_\star = 1 + i\hat{\Lambda} + \frac{1}{2}i\hat{\Lambda}\star i\hat{\Lambda} +$...

 From now on,  we include the coupling constant
$q$  between fields $\Phi$ and $A_\mu$, i.e. the charge of $\Phi$,  into $A_\mu$, so that the redefinition $A_\mu \longrightarrow qA_\mu$ is implicitly understood. Then we use the  Seiberg-Witten (SW)-map \cite{seiberg-witten, seiberg-witten2} in order  to express  NC fields $ \hat{\Phi}$, $\hat{A}_\mu$ and $\hat{F}_{\mu\nu}$  as functions of corresponding commutative fields
and the deformation parameter $a$. SW-map assumes an expansion  in orders of the deformation parameter  and this expansion is known to all orders for an arbitrary Abelian twist deformation \cite{PLSWGeneral}, of which the twist (\ref{AngTwist0Phi})  is only one example.
For  the twist operator (\ref{AngTwist0Phi}), SW-map gives rise to the following expansions for the fields:
\begin{eqnarray}
\hat{\Phi} &=& \Phi -\frac{1}{4}\theta^{\rho\sigma}A_\rho(\partial_\sigma\Phi + (\partial_\sigma - i A_\sigma)
\Phi), \label{HatPhi}\\
\hat{A}_\mu &=& A_\mu -\frac{1}{2}\theta^{\rho\sigma}A_\rho(\partial_\sigma A_{\mu} +
F_{\sigma\mu}), \label{HatA}\\
\hat{F}_{\mu\nu} &=& F_{\mu\nu} - \theta^{\rho\sigma}A_{\rho} \partial_\sigma F_{\mu\nu} +\theta^{\rho\sigma}F_{\rho\mu}F_{\sigma\nu}. \label{HatFmunu}
\end{eqnarray}


Using the SW-map solutions and expanding the $\star$-products in (\ref{SPhi}) and (\ref{SA}) we
find the action up to first order in the deformation parameter $a$,
\begin{eqnarray}
S &=& \int
\d^4x\sqrt{-g}\,
\bigg( -\frac{1}{4q^2}g^{\mu\rho}g^{\nu\sigma}F_{\mu\nu}F_{\rho\sigma}
+ g^{\mu\nu} \Big[ (\partial_{\mu} - iA_{\mu}) \Phi^+  \Big]  (\partial_{\nu} - iA_{\nu})  \Phi -\mu^2\Phi^+\Phi \nonumber\\
&&
+\frac{1}{8q^2}g^{\mu\rho}g^{\nu\sigma}\theta^{\alpha\beta}(F_{\alpha\beta}F_{\mu\nu}F_{\rho\sigma}
-4F_{\mu\alpha}F_{\nu\beta}F_{\rho\sigma}) +\frac{\mu^2}{2}\theta^{\alpha\beta}F_{\alpha\beta}\Phi^+\Phi \label{SExp}\\
&& + \frac{\theta^{\alpha\beta}}{2}g^{\mu\nu}\Big( -\frac{1}{2}   \Big[ (\partial_{\mu} - iA_{\mu}) \Phi^+  \Big]   F_{\alpha\beta}
   (\partial_{\nu} - iA_{\nu})  \Phi
+  \Big[ (\partial_{\mu} - iA_{\mu}) \Phi^+  \Big]    F_{\alpha\nu}   
  (\partial_{\beta} - iA_{\beta})   \Phi \nonumber  \\
 &&  +     \Big[ (\partial_{\beta} - iA_{\beta}) \Phi^+  \Big]   F_{\alpha\mu}     (\partial_{\nu} - iA_{\nu})   \Phi  \Big)   \bigg).  \nonumber
\end{eqnarray}

 
Since the gauge field is the electromagnetic field generated by the RN black hole charge and RN black hole is non-rotating,  only the time component $A_t$ of the gauge field is nonvanishing.
The corresponding field strength will consequently have $F_{rt}=-F_{tr} $ as the  only  components different from zero,
\begin{equation}  \label{A0Fr0}
A_t = -\frac{qQ}{r}, \quad    F_{rt} = \frac{qQ}{r^2}.
\end{equation}
We also note that  the only components of 
$\theta^{\alpha\beta}$ that are different from zero are
$\theta^{t\varphi}=
-\theta^{\varphi t}=a$. Using these and varying the action (\ref{SExp}) with respect to the field $\Phi^+$ and using the above, we can write the equation of motion as
\begin{equation}  \label{extendedKG}
\Big( \frac{1}{f}\partial^2_t -\Delta + (1-f)\partial_r^2 
+\frac{2MG}{r^2}\partial_r + 2iqQ\frac{1}{rf}\partial_t -\frac{q^2Q^2}{r^2f} -\mu^2 \Big) \Phi  +\frac{aqQ}{r^3}
\Big( (\frac{MG}{r}-\frac{GQ^2}{r^2})\partial_\phi
+ rf\partial_r\partial_\phi \Big) \Phi =0 
\end{equation}
which after the separation of variables gives the radial equation \eqref{NCRNequation}.



\begin{thebibliography}{99}

\bibitem{Bekenstein}
J.~D.~Bekenstein,
``Black holes and the second law,''
Lett. Nuovo Cim. \textbf{4} (1972), 737-740
doi:10.1007/BF02757029

\bibitem{Hawking}
S.~W.~Hawking,
``Particle Creation by Black Holes,''
Commun. Math. Phys. \textbf{43} (1975), 199-220
[erratum: Commun. Math. Phys. \textbf{46} (1976), 206]
doi:10.1007/BF02345020

\bibitem{thorne-zurek}  K. S. Thorne  and  W. H. Zurek, ``Statistical Mechanical Origin of the Entropy of a Rotating, Charged Black Hole", Phys. Rev. Lett.  {\bf 54}, 2171  (1985).

\bibitem{thooft} G.~'t Hooft,``On the Quantum Structure of a Black Hole,''
Nucl. Phys. B \textbf{256} (1985), 727-745
doi:10.1016/0550-3213(85)90418-3 

\bibitem{thooft1}
G.~'t Hooft,
``The Scattering matrix approach for the quantum black hole: An Overview,''
Int. J. Mod. Phys. A \textbf{11} (1996), 4623-4688
doi:10.1142/S0217751X96002145
[arXiv:gr-qc/9607022 [gr-qc]]

\bibitem{vafa} A. Strominger and C. Vafa, ``Microscopic Origin of the Bekenstein-Hawking Entropy", Phys. Lett. {\bf B 379}, 99 (1996).

\bibitem{ashtekar} A. Ashtekar, J. Baez, A. Corichi and K. Krasnov, ``Quantum Geometry and Black Hole Entropy", Phys. Rev. Lett. {\bf 80}, 904 (1998). 

\bibitem{strominger} A. Strominger, ``Black Hole Entropy from Near-Horizon Microstates", JHEP {\bf 9802}, 009 (1998).

\bibitem{carlip} S. Carlip, ``Entropy from Conformal Field Theory at Killing Horizons", 
Class. Quant. Grav. {\bf 16}, 3327 (1999).

\bibitem{kaul} Romesh K. Kaul and Parthasarathi Majumdar, ``Logarithmic Correction to the Bekenstein-Hawking Entropy", Phys. Rev. Lett. {\bf 84}, 5255 (2000).

\bibitem{Fursaev:1994ea}
D.~V.~Fursaev and S.~N.~Solodukhin,
``On one loop renormalization of black hole entropy,''
Phys. Lett. B \textbf{365} (1996), 51-55
doi:10.1016/0370-2693(95)01290-7
[arXiv:hep-th/9412020 [hep-th]].

\bibitem{carlip1} S. Carlip, ``Logarithmic corrections to black hole entropy, from the Cardy formula", Class. Quantum Grav. {\bf 17}, 4175 (2000).

\bibitem{ksgsen} Kumar S. Gupta and Siddhartha Sen, ``Further evidence for the conformal structure of a Schwarzschild black hole in an algebraic approach", Phys. Lett. {\bf B 526}, 121 (2002).

\bibitem{asen1} Ashoke Sen, ``Logarithmic Corrections to N=2 Black Hole Entropy: An Infrared Window into the Microstates", Gen. Rel. and Grav. {\bf 44}, 1207 (2012).

\bibitem{asen2} Ashoke Sen, ``Logarithmic Corrections to Schwarzschild and Other Non-extremal Black Hole Entropy in Different Dimensions", JHEP {\bf 04}, 156 (2013).

\bibitem{sud} Sudipta Mukherji and Shesansu Sekhar Pal, ``Logarithmic corrections to black hole entropy and AdS / CFT correspondence", JHEP {\bf 05} 026 (2002).

\bibitem{solo} Sergey N.Solodukhin, ``Logarithmic terms in entropy of Schwarzschild black holes in higher loops", Phys. Lett. {bf B 802}, 135235 (2020).

\bibitem{solo1}
S.~N.~Solodukhin,
``Entanglement entropy of black holes,''
Living Rev. Rel. \textbf{14} (2011), 8
doi:10.12942/lrr-2011-8
[arXiv:1104.3712 [hep-th]].

\bibitem{xiao} Y.~Xiao and Y.~Tian,
``Logarithmic correction to black hole entropy from the nonlocality of quantum gravity,''
Phys. Rev. D \textbf{105} (2022) no.4, 044013
[arXiv:2104.14902 [gr-qc]].

\bibitem{Calmet:2021lny}
X.~Calmet and F.~Kuipers,
``Quantum gravitational corrections to the entropy of a Schwarzschild black hole,''
Phys. Rev. D \textbf{104} (2021) no.6, 066012
[arXiv:2108.06824 [hep-th]].

\bibitem{bombelli-sorkin}  L. Bombelli, R. K. Koul, J. Lee and R. D. Sorkin, Phys. Rev. D 34 (1986) 373.

\bibitem{srednicki} M. Srednicki, Phys. Rev. Lett. 71 (1993) 666.

\bibitem{Bodendorfer:2013wga}
N.~Bodendorfer and Y.~Neiman,
``Wald entropy formula and loop quantum gravity,''
Phys. Rev. D \textbf{90} (2014) no.8, 084054
doi:10.1103/PhysRevD.90.084054
[arXiv:1304.3025 [gr-qc]].

\bibitem{das} ``Black hole entropy from entanglement: A review", Saurya Das, S. Shankaranarayanan and Sourav Sur, Horizons in World Physics, Volume 268 (2009) Editors: Michael Everett and Louis Pedroza, [arXiv:0806.0402 [gr-qc] ].

\bibitem{dfr1} S.~Doplicher, K.~Fredenhagen and J.~E.~Roberts,
``Space-time quantization induced by classical gravity,''
Phys. Lett. B \textbf{331} (1994), 39-44
doi:10.1016/0370-2693(94)90940-7

\bibitem{dfr2} S.~Doplicher, K.~Fredenhagen and J.~E.~Roberts,
``The Quantum structure of space-time at the Planck scale and quantum fields,''
Commun. Math. Phys. \textbf{172} (1995), 187-220
doi:10.1007/BF02104515
[arXiv:hep-th/0303037 [hep-th]].

\bibitem{szabo}
R.~J.~Szabo,
``Quantum field theory on noncommutative spaces,''
Phys. Rept. \textbf{378} (2003), 207-299
[arXiv:hep-th/0109162 [hep-th]].

\bibitem{Chaichian:2004za}
M.~Chaichian, P.~P.~Kulish, K.~Nishijima and A.~Tureanu,
``On a Lorentz-invariant interpretation of noncommutative space-time and its implications on noncommutative QFT,''
Phys. Lett. B \textbf{604} (2004), 98-102
[arXiv:hep-th/0408069 [hep-th]].

\bibitem{Aschieri1}
P.~Aschieri, F.~Lizzi and P.~Vitale,
``Twisting all the way: From Classical Mechanics to Quantum Fields,''
Phys. Rev. D \textbf{77} (2008), 025037
[arXiv:0708.3002 [hep-th]].
\bibitem{Aschieri2}
P.~Aschieri,
``Lectures on Hopf Algebras, Quantum Groups and Twists,''
[arXiv:hep-th/0703013 [hep-th]].

\bibitem{sub}
S.~Sarkar, S.~Shankaranarayanan and L.~Sriramkumar,
``Sub-leading contributions to the black hole entropy in the brick wall approach,''
Phys. Rev. D \textbf{78} (2008), 024003
doi:10.1103/PhysRevD.78.024003
[arXiv:0710.2013 [gr-qc]].

\bibitem{nicolini} Jonas R. Mureika and Piero Nicolini, ``Aspects of noncommutative (1+1)-dimensional black holes", 	Phys. Rev. {\bf D 84}, 044020 (2011)

\bibitem{gup} M. A. Anacleto, F. A. Brito, S. S. Cruz and E. Passos, ``Noncommutative correction to the entropy of Schwarzschild black hole with GUP", Int. J. Mod. Phys. {\bf A 36}, 2150028 (2021). 

\bibitem{Banerjee:2008gc}
R.~Banerjee, B.~R.~Majhi and S.~Samanta,
``Noncommutative Black Hole Thermodynamics,''
Phys. Rev. D \textbf{77} (2008), 124035
doi:10.1103/PhysRevD.77.124035
[arXiv:0801.3583 [hep-th]].

\bibitem{Lizzi} 
M.~Dimitrijevic Ciric, N.~Konjik, M.~A.~Kurkov, F.~Lizzi and P.~Vitale,
``Noncommutative field theory from angular twist,''
Phys. Rev. D \textbf{98} (2018) no.8, 085011
doi:10.1103/PhysRevD.98.085011
[arXiv:1806.06678 [hep-th]].

\bibitem{andelo}
M.~D.~\'Ciri\'c, N.~Konjik and A.~Samsarov,
``Noncommutative scalar quasinormal modes of the Reissner\textendash{}Nordstr\"om black hole,''
Class. Quant. Grav. \textbf{35} (2018) no.17, 175005
doi:10.1088/1361-6382/aad201
[arXiv:1708.04066 [hep-th]].

\bibitem{DimitrijevicCiric:2019hqq}
M.~Dimitrijevi\'c \'Ciri\'c, N.~Konjik and A.~Samsarov,
``Noncommutative scalar field in the nonextremal Reissner-Nordstr\"om background: Quasinormal mode spectrum,''
Phys. Rev. D \textbf{101} (2020) no.11, 116009
doi:10.1103/PhysRevD.101.116009
[arXiv:1904.04053 [hep-th]].

\bibitem{Ciric:2019urb}
M.~D.~\'Ciri\'c, N.~Konjik and A.~Samsarov,
``Search for footprints of quantum spacetime in black hole QNM spectrum,''
[arXiv:1910.13342 [hep-th]].

\bibitem{Konjik:2020fle}
N.~Konjik, M.~D.~\'Ciri\'c and A.~Samsarov,
``Noncommutative field theory from an angular twist,''
PoS \textbf{CORFU2019} (2020), 231

\bibitem{C1} M. Chaichian, A. Demichev, P. Presnajder and A. Tureanu, Eur. Phys. J. {\bf C 20} (2001) 767.

\bibitem{C2} M. Chaichian, A. Demichev, P. Presnajder and A. Tureanu, Phys. Lett. {\bf B 515} (2001) 426.

\bibitem{L1} J. Lukierski, A. Nowicki, H. Ruegg and V. N. Tolstoy, Phys. Lett.  {\bf B 264} (1991) 331.

\bibitem{L2} J. Lukierski, A. Nowicki and  H. Ruegg, Phys. Lett.  {\bf B 293} (1992) 344.

\bibitem{L3} J. Lukierski and H. Ruegg, Phys. Lett.  {\bf B 329} (1994) 189. 

\bibitem{L4} J. Lukierski, H. Ruegg and W. J. Zakrzewski, Ann. Phys. {\bf 243} (1995) 90.

\bibitem{brian1} B.P. Dolan, Kumar S. Gupta and A. Stern, Class. Quant. Grav. {\bf 24} (2007) 1647. 

\bibitem{brian2} B.P. Dolan, Kumar S. Gupta and A. Stern, J. Phys. Conf. Ser. {\bf 174} (2009) 012023.

\bibitem{schupp} P. Schupp and S. Solodukhin, arXiv:0906.2724 [hep-th].

\bibitem{ksgbh} Kumar S. Gupta, Stjepan Meljanac and Andjelo Samsarov, ``Quantum statistics and noncommutative black holes", Phys. Rev. {\bf D 85}, 045029 (2012). 

\bibitem{ohl} T. Ohl and A. Schenkel, JHEP {\bf 0910} (2009) 052. 

\bibitem{Govindarajan:2009wt}
T.~R.~Govindarajan, K.~S.~Gupta, E.~Harikumar, S.~Meljanac and D.~Meljanac,
``Deformed Oscillator Algebras and QFT in kappa-Minkowski Spacetime,''
Phys. Rev. D \textbf{80} (2009), 025014
doi:10.1103/PhysRevD.80.025014
[arXiv:0903.2355 [hep-th]].

\bibitem{Dimitrijevic:2011jg}
M.~Dimitrijevic and L.~Jonke,
``A Twisted look on kappa-Minkowski: U(1) gauge theory,''
JHEP \textbf{12} (2011), 080
doi:10.1007/JHEP12(2011)080
[arXiv:1107.3475 [hep-th]].

\bibitem{dim2} M. Dimitrijevi\'{c}, L. Jonke and A. Pachol,
SIGMA 10, 063 (2014), [arXiv:1403.1857].



\bibitem{22}
P. Kulish, F. Lizzi and J. Wess Noncommutative
spacetimes: Symmetries in noncommutative geometry and field theory, Lecture
notes in physics 774, Springer (2009).

\bibitem{23} P. Aschieri and L. Castellani, Noncommutative D = 4 gravity coupled to fermions
JHEP, 0906, 086 (2009), [arXiv:0902.3823].

\bibitem{seiberg-witten} 
S. Schraml, P. Schupp and J. Wess, Construction of non-Abelian gauge theories on noncommutative spaces, Eur. Phys. J. C21, 383 (2001),
[hep-th/0104153].

\bibitem{seiberg-witten2} N. Seiberg and E. Witten, String theory and noncommutative geometry, JHEP
09, 032 (1999), [hep-th/9908142].

\bibitem{PLSWGeneral}
P. Aschieri and L. Castellani, {\it Noncommutative gravity 
coupled to fermions: second order expansion via Seiberg-Witten map}, JHEP {\bf
1207} 184 (2012), [arXiv:1111.4822].






\bibitem{Demers:1995dq}
J.~G.~Demers, R.~Lafrance and R.~C.~Myers,
``Black hole entropy without brick walls,''
Phys. Rev. D \textbf{52} (1995), 2245-2253
doi:10.1103/PhysRevD.52.2245
[arXiv:gr-qc/9503003 [gr-qc]].




\bibitem{luk1} J. Lukierski, H. Ruegg, V. N. Tolstoi and A. Nowicki, Twisted classical Poincare
algebras, J. Phys. A 27 (1994) 2389, [hep-th/9312068].

\bibitem{luk2} J. Lukierski and M. Woronowicz, New Lie-algebraic and quadratic deformations
of Minkowski space from twisted Poincare symmetries, Phys. Lett. B 633 (2006)
116, [hep-th/0508083].

\bibitem{Meljanac:2017oek}
D.~Meljanac, S.~Meljanac, D.~Pikuti\'c and K.~S.~Gupta,
Phys. Rev. D \textbf{96} (2017) no.10, 105008
doi:10.1103/PhysRevD.96.105008
[arXiv:1703.09511 [hep-th]].


\bibitem{inside} W.~B.~Liu,
``Reissner-Nordstrom black hole entropy inside and outside the brick wall,''
Chin. Phys. Lett. \textbf{20} (2003), 440-443
doi:10.1088/0256-307X/20/3/337

\bibitem{Aschieri:2009qh}
P.~Aschieri and L.~Castellani,
``Noncommutative Gravity Solutions,''
J. Geom. Phys. \textbf{60} (2010), 375-393
doi:10.1016/j.geomphys.2009.11.009
[arXiv:0906.2774 [hep-th]].

\bibitem{Schenkel:2011biz}
A.~Schenkel,
``Noncommutative Gravity and Quantum Field Theory on Noncommutative Curved Spacetimes,'' PhD thesis
[arXiv:1210.1115 [math-ph]].

\bibitem{Witten:2021unn}
E.~Witten,
``Gravity and the Crossed Product,''
[arXiv:2112.12828 [hep-th]].

\bibitem{witten1} E. Witten, {\it Why Does Quantum Field Theory In Curved Spacetime Make Sense? And What Happens To The Algebra of Observables In The Thermodynamic Limit?}, [arXiv:2112.11614 [hep-th]]

\bibitem{witten-longo} V.~Chandrasekaran, R.~Longo, G.~Penington and E.~Witten,
``An Algebra of Observables for de Sitter Space,''
[arXiv:2206.10780 [hep-th]].



\bibitem{Gupta:2013ata}
K.~S.~Gupta, E.~Harikumar, T.~Juric, S.~Meljanac and A.~Samsarov,
``Effects of Noncommutativity on the Black Hole Entropy,''
Adv. High Energy Phys. \textbf{2014} (2014), 139172
doi:10.1155/2014/139172
[arXiv:1312.5100 [hep-th]].

\bibitem{Juric:2016zey}
T.~Juri\'c and A.~Samsarov,
``Entanglement entropy renormalization for the noncommutative scalar field coupled to classical BTZ geometry,''
Phys. Rev. D \textbf{93} (2016) no.10, 104033
doi:10.1103/PhysRevD.93.104033
[arXiv:1602.01488 [hep-th]].

\bibitem{Susskind:1994sm}
L.~Susskind and J.~Uglum,
``Black hole entropy in canonical quantum gravity and superstring theory,''
Phys. Rev. D \textbf{50} (1994), 2700-2711
doi:10.1103/PhysRevD.50.2700
[arXiv:hep-th/9401070 [hep-th]].

\bibitem{Fursaev:1994in}
D.~V.~Fursaev,
``Spectral geometry and one loop divergences on manifolds with conical singularities,''
Phys. Lett. B \textbf{334} (1994), 53-60
doi:10.1016/0370-2693(94)90590-8
[arXiv:hep-th/9405143 [hep-th]].










\end{thebibliography}
\end{document}